\documentclass{jfm} 
\usepackage{afterpage} 
\usepackage{amsmath} 
\usepackage{amssymb} 
\usepackage{booktabs}
\usepackage{dcolumn}
\usepackage{graphicx}
\usepackage{natbib}
\usepackage{rotating}
\usepackage{setspace}
\usepackage{upmath}
\usepackage{afterpage}
\usepackage{mathtools}
\usepackage{bm}
\usepackage{float}
\usepackage{epstopdf}
\usepackage{subcaption}
\usepackage{pstricks}
\usepackage{xcolor}

\newcommand{\Sc}{\ensuremath{\mathit{Sc}}}
\newcommand{\ReL}{\ensuremath{\mathit{Re_L}}}

\newcommand{\Pey}{\ensuremath{\mathit{Pe_\lambda}}}

\newcommand{\Lk}{\ensuremath{L_k}}
\newcommand{\Lc}{\ensuremath{L_c}}
\newcommand{\Lb}{\ensuremath{L_b}}
\newcommand{\Lr}{\ensuremath{L_r}}
\newcommand{\Rey}{\ensuremath{\mathit{Re_\lambda}}}

\newcommand{\Xpsi}{\bm{X}_{\psi}} 

\DeclareMathSymbol{\varchi}{\mathord}{letters}{88}

\begin{document}
\setcounter{tocdepth}{10}  
\setcounter{secnumdepth}{4}

\title[Area of Scalar Isosurfaces]{Area of Scalar Isosurfaces in
	Homogeneous Isotropic Turbulence as a Function of Reynolds and Schmidt
	Numbers}

\author[K.~Prashant and S.~M.~de Bruyn Kops]%
{Kedar Prashant Shete and Stephen M. de Bruyn Kops}
\affiliation{Department of Mechanical and Industrial Engineering, \\University
	of Massachusetts Amherst, Amherst,
	MA 01003-9284, USA}

\maketitle

\begin{abstract}
A fundamental effect of fluid turbulence is turbulent mixing, which results in
the stretching and wrinkling of scalar isosurfaces.  Thus, the area of
isosurfaces is of interest in understanding turbulence in general with
specific applications in, e.g., combustion and the identification of
turbulent/non-turbulent interfaces.  We report measurements of isosurface
areas in 28 direct numerical simulations (DNSs) of homogeneous isotropic
turbulence with a mean scalar gradient resolved on up to $14256^3$ grid points
with Taylor Reynolds number $\Rey$ ranging from 24 to 633 and Schmidt number
$Sc$ ranging from 0.1 to 7.  More precisely, we measure layers with very small
but finite thickness.  The continuous equation we evaluate converges exactly
to the area in the limit of zero layer thickness.  We demonstrate a method for
numerically integrating this equation that, for a test case with an analytical
solution, converges linearly towards the exact solution with decreasing layer
width.  By applying the technique to DNS data and testing for convergence with
resolution of the simulations, we verify the resolution requirements for DNS
recently proposed by \citet{yeung18}.  We conclude that isosurface areas
scale with the square root of the Taylor P\'eclet number $\Pey$ between
approximately 50 and 4429 with some departure from power law scaling evident
for $2.4 < $\Pey$ < 50$.  No independent effect of either $\Rey$ or $Sc$ is
observed.  The excellent scaling of area with $\Pey^{1/2}$ occurs even though
the probability density function (p.d.f.) of the scalar gradient is very close to
exponential for $\Rey=98$ but approximately lognormal when $\Rey=633$.
\end{abstract}

\tableofcontents
\clearpage
\section{Introduction}
The area of an isosurface in a turbulent flow reflects the instantaneous
balance between advection that stretches and compresses an isosurface,
diascalar diffusion that smooths an isosurface, and possibly the effects of
other phenomena such as chemical reaction.  This is true whether the scalar is
passive or active.  Thus an isosurface area is a composite measure that is
of fundamental interest because of what it can tell us about fluid turbulence.
Our motivation here is the general question of how an isosurface area 
changes with Reynolds and Schmidt numbers, but let us first
consider several applications requiring knowledge of an isosurface area.

A principal motivation for understanding isosurfaces is combustion modelling.
In both premixed and non-premixed combustion, the rate of chemical reaction
depends on the volume in which the species concentrations and temperature are
suitable for reaction. It also typically depends on the mass flux of one
scalar through an isosurface of another.  Calculating either of these
quantities is a generalisation of the problem of finding the area of one or
more isosurfaces \citep{pope88,kim07}.  It is, however, impossible to
resolve isosurfaces in simulations that are practical for the
design of combustion devices and so these quantities must be parameterised.
One avenue of parameterisation is via the isosurface area to volume ratio as a
function of flow parameters including the Reynolds, Schmidt, and Damk\"oholer
numbers, etc. \citet{peters00} explored the importance of calculating
iso-scalar surfaces and their role in combustion modelling. One of the earlier
attempts of modelling combustion using iso-scalar surface statistics is given
in \citet{swaminathan06}. \citet[pp. 41-60]{bray11} discuss the equation for
isosurface area and its uses in modelling premixed combustion. \citet{bray16}
studied laminar flamelets in premixed combustion. A more recent example of
combustion analysis based on an isosurface area is that of
\citet{Chaudhuri2017} who studied the geometric flame thickness of a reacting
jet. \citet{zheng17} studied the propagation of
scalar isosurfaces in isotropic turbulence and formulated a model for
estimating isosurface area changes with time in order to model a planar
premixed flame front. Motivated by the modelling of premixed combustion, \citet{dopazo18}
considers the propagation of scalar isosurfaces and also provides an equation
for the time rate of change of the infinitesimal isosurface area
\citep[][equation 25]{dopazo18}. 

Applications for isosurface area calculations outside the field of combustion
include the size of the entrainment area in a turbulent jet
\citep{ricou61,delichatsios87} and the area of the turbulent/non-turbulent
interface \citep{corrsin55,taveira14,watanabe16}.  \citet{schumacher05a}
considered the relationships between the geometrical and statistical
properties of a scalar field. In general, these quantities are important for
understanding and modelling mixing in environmental and technological
flows. \citet{watanabe16} consider internal wave energy transported through an
isosurface of enstrophy in a stably stratified wake.

Methods for calculating an isosurface area can be divided into two types,
which we refer to as ``geometric methods'' and ``integral methods''.
Geometric methods involve multiple steps, each potentially involving numerical
or mathematical approximations.  First, discrete points on an isosurface are
located by a process that usually requires some methods of interpolation and
root finding.  Next, the points are connected by surface ``patches'' to
approximate the surface.  In general, these surface patches can be curved,
although typically planar polygons are used thereby introducing the assumption
that the surface is locally planar at the resolution at which it is sampled.
Finally, the areas of surface patches are integrated, which introduces another
approximation, with the most common approach being to sum the areas of the
polygons to yield the area of the surface.  Methods of this type include
marching cubes, marching tetrahedra, and surface nets; a review of the
geometrical methods can be found in \citet{Patera2004}. Geometric methods are
attractive when it is desirable to visualise the surface in addition to
estimating the area, and indeed, Paraview, {a very popular} open source data
visualisation software {used in computational fluid dynamics}, uses the
marching cubes algorithm for constructing isosurfaces \citep[][Chapter
  6]{schroeder06}.

Drawbacks to geometric methods include that they do not always converge to the
true surface area as the number of discrete points on the surface is
increased, topological ambiguities can result in problems connecting those
points to form a surface, and the algorithms are computationally complex
\citep{liu10}.  \citet{newman06} conducted a review of the marching cubes
algorithm and noted that there are many variations within it and a number of
``spin-off'' methods to improve certain aspects of the basic algorithm.  In
particular, considerable research has been done into how to resolve the
potential ambiguities in connecting discrete points to form a surface.
\citet{pope89} concluded that level scalar surfaces
can be highly wrinkled and disjoint in turbulent reacting flows and a direct
geometrical representation may not be feasible.  \citet{lewiner12} concluded
that there can be connectivity issues even with relatively simple convex
isosurface geometries.

While geometric methods are based on the intuitive approach of finding the
surface and measuring its area, they are complicated by a variety of
approximations required to implement the approach.  Integral methods, in
contrast, begin with an exact equation for the surface, which may not be
intuitive.  Our approach, detailed in \S\ref{sec:arearatio}, is based on
Federer's coarea formula which relates the surface area to the infinitesimally
thin volume that contains it \citep{federer59}.  Given this equation, the
problem of finding an isosurface area reduces to that of integrating the
equation numerically.  Discussion of our specific approach to doing this is
deferred to \S\ref{sec:arearatio}, but we note that the general approach has
been reported in the literature \citep{yurtoglu18}.  They conducted
convergence tests for an integral approach to calculating the volume of a
torus.  They also provide a proof, based on that of \citet{resnikoff12}, for
the surface integral of a function over a level scalar set \citep[][equation
  10]{yurtoglu18}, which when discretised for an infinitesimal scalar range
goes over to the equation for an isosurface area in
\S\ref{sec:arearatio}. Isosurface statistics summed over the scalar
domain have also been a point of interest \citep{scheidegger08} and the area
equation in \S\ref{sec:arearatio} agrees with their result upon integration
over the entire range of scalar values.


An additional consideration when choosing a method for calculating
isosurfaces is advances in computer architecture.  As early as 1988,
\citet{payne88} noted that massively parallel computers had enabled the
calculation of flow fields that were previously intractable in simulations.
Since then, the importance of parallel algorithms have increased dramatically,
particularly since the breakdown of Dennard scaling in approximately 2006
\citep{dennard74,bohr07} and the resulting requirement that, for an algorithm
to run faster, it must exploit additional parallelism on a single computer
processor and across multiple processors.  \citet{raase15}, for example,
reported the utility of many-core processors for fluids simulations.
\citet{Kolla2018}, though, report that geometric methods for finding
isosurface areas are difficult to parallelize.  
Monte-Carlo techniques suitable for evaluating integral
methods for computing isosurface areas are well-suited for parallelisation
\citep{dimov07}, and this is the approach we take, as discussed in
\S\ref{sec:simple}.  \citet{yurtoglu18} present an approach specifically
designed to avoid the stochastic nature and slow convergence of Monte-Carlo
methods while being highly scalable on general purpose graphics processors.

The specific subjects of this paper are (1) the demonstration of a integral
method for computing isosurfaces and (2) its application to direct numerical
simulation (DNS) data to understand how an isosurface area varies with
Reynolds, Schmidt, and P\'eclet numbers.  The integral method is presented in
\S\ref{sec:arearatio} and verified with a simple case in \S\ref{sec:simple}.
The DNS are presented in \S\ref{sec:dns} followed by analysis in
\S\ref{sec:numerical} and \S\ref{sec:results}. Some conclusions are drawn in \S\ref{sec:conclusions}.

\section{An integral expression for an isosurface area }
\label{sec:arearatio}
Let us first define some notation. Consider a scalar field $\phi(\vec{x})$ to
be a function that maps $\vec{x}$ to a scalar $\phi$. The volume in which we
want an isosurface area is $V$, the set of all possible position vectors in
$V$ is $\bf X$, and the set of all values of $\phi$ in $V$ is $\bm \phi$. The
subset of $\bm X$ at which $\phi=\psi$ is $\bm X_\psi$.  Assuming that an
isosurface exists with isovalue $\psi$ then its area is $A(\psi)$ and the
differential area associated with each point in $\bm{X}_\psi $ is $ \hat{n} dA
$ with $\hat{n} = \nabla{\phi}/\left| \nabla{\phi}\right|$. The volume
between isosurfaces with $\phi=\psi$ and $\phi=\psi+ d\phi$ corresponds to the
set of all the position vectors between $ \bm{X}_\psi $ and $
\bm{X}_{\psi+d\phi} $.  Consider a position vector $\vec{x} = \vec{\xi}$ such
that $\vec{\xi} \in \bm{X}_\psi$.  For every $\vec{\xi}$, since the gradient
$\nabla\phi$ is defined at every point, there exists a differential vector
$d\vec{x} = \hat{n}dx$ such that
\begin{equation}
\vec{x} = \vec{\xi} + d\vec{x} \in \bm{X}_{\psi + d\phi} \ .
\label{eq:defdx}
\end{equation}
Hence, a volume element at $\vec{\xi}$ corresponding to the volume between ensembles $\bm{X}_{\psi}$ and $\bm{X}_{\psi + d\phi}$ is given by $\hat{n} dA \cdot \hat{n} dx$. 
The volume of this region is given by integrating over $\bm{X}_{\psi}$:
\begin{equation}
dV = \int_{\bm{X}_\psi}^{} \hat{n} dA \cdot \hat{n} dx  \ .
\label{eq:diffvolume}
\end{equation}
Now consider a finite thickness of a layer where the infinitesimal width
$d\phi$ is replaced by a finite width $\Delta \phi$. The volume $dV$ given in
\eqref{eq:diffvolume} integrated over the region between $\Xpsi$ and
$\bm{X}_{\psi + \Delta \phi}$ gives us the volume of the finite layer between
$\Xpsi$ and $\bm{X}_{\psi + \Delta \phi}$, which is represented by
$V(\psi,\Delta \phi)$.  Now consider Federer's coarea formula
\begin{equation}
	\label{eq:federer_coarea}
	\int_{\psi}^{\psi + \Delta \phi} \int_{\Xpsi} q(\vec{x}) dA\ d\phi = \int_{V(\psi,\Delta \phi)} q(\vec{x}) |\nabla \phi| dV
\end{equation}
which relates integrals over surfaces to integrals over volumes
\citep{federer59}.  This particular form of the coarea formula is from
\citet[][equation 3]{scheidegger08} where $q(\vec{x})$ is any scalar function
defined over the domain of $\phi$ {and $\phi(\vec{x})$ is a Lipschitz
  continuous function. Lipschitz continuity implies that $|\nabla \phi|$ is
  bounded, which is true in most data generated by numerical simulation
  because finite derivatives are required to solve the governing equations.
  Note that $V(\psi,\Delta \phi)$ can be any volume that equals or is a subset
  of the domain of $\phi$.  The term on the right hand side can be written as
  the ensemble average of $q(\vec{x})|\nabla \phi|$ over $V(\psi,\Delta \phi)$
  times the volume $V(\psi,\Delta \phi)$. Upon setting $q(\vec{x})=1$ in
  \eqref{eq:federer_coarea}, the inner integral becomes the area of
  $\Xpsi$. Dividing both sides by $V$, \eqref{eq:federer_coarea} simplifies to
\begin{equation}
\label{eq:federer_coarea_integral}
\frac{1}{V}\int_{\psi}^{\psi + \Delta \phi} A(\phi) d\phi =
\frac{\langle|\nabla\phi|\rangle_{V(\psi,\Delta \phi)} V(\psi,\Delta \phi)}{V}
\ ,
\end{equation}
which is an exact equation as it is a direct consequence of Federer's coarea
formula.  Approximating the outer integral using the rectangle rule yields the discrete equation 
\begin{equation}
\label{eq:coarea_1}
\frac{A(\psi)}{V} \Delta \phi \approxeq \frac{\langle|\nabla\phi|\rangle_{V(\psi,\Delta \phi)} V(\psi,\Delta \phi)}{V}
\end{equation}
which we can evaluate numerically and which goes over to
\eqref{eq:federer_coarea_integral} in the limit of $\Delta\phi \rightarrow 0$.

The volume $V(\psi,\Delta \phi)$ in \eqref{eq:coarea_1} represents all
possible position vectors $\vec{x}$ such that $\phi(\vec{x}) \in [\psi,\psi +
  \Delta \phi)$. Since $V$ is the domain of $\phi$, it represents all possible
  position vectors that correspond to the range of $\phi$. Dividing both sides by $\Delta \phi$,
  \eqref{eq:coarea_1} can be simplified to
  \begin{equation}
\label{eq:coarea_arearatio}
\frac{A(\psi)}{V} \approxeq \langle|\nabla\phi|\rangle_{V(\psi,\Delta \phi)} \frac{V(\psi,\Delta \phi)}{V\ \Delta \phi}
\end{equation}
which is a first order approximation with respect to $\Delta \phi$ for the
area to volume ratio of an isosurface given by $\Xpsi$, and it converges to
the exact area ratio as $\Delta \phi$ is reduced.  This is the equation we
evaluate to compute areas in the remainder of the paper.

\section{A simple example}
\label{sec:simple}
To illustrate the evaluation of isosurface area via \eqref{eq:coarea_arearatio},
we consider the scalar field 
\begin{equation}
\phi(\vec{x})=x^2 + y^2 + 2z
\end{equation}
in the volume $x\in[0,2],y\in[0,2],z\in[0, 2]$.  In this test case $\phi$ is
known exactly so that the area of any isosurface and the mean gradient on the
surface are known analytically. We use the exact area as a reference against
which to compare numerical solutions of \eqref{eq:coarea_arearatio}.

Numerically evaluating \eqref{eq:coarea_arearatio} via a Monte-Carlo method
reduces to sampling $\phi$ and accumulating statistics for $\nabla\phi$.  Looking ahead
in this paper, we note that the DNS database we use is produced by a
pseudo-spectral simulation so that, given $\phi$ from the simulation, no
numerical error is introduced in computing either $\phi$ or $\nabla\phi$ at an
arbitrary location $\vec{x}$; why this is so is explained in
\S\ref{sec:numerical}.  Therefore, the uncertainty in evaluating
\eqref{eq:coarea_arearatio} is dependent on how $\phi$ is sampled, the number
of times it is sampled, and the layer thickness $\Delta \phi$.

To evaluate \eqref{eq:coarea_arearatio} for this test case, $\phi(\vec{x})$ is
sampled at random locations chosen using a pseudo-random number generator with
the term ``pseudo-random'' referring to how sequences are computed on a
typical digital computer.  There are advantages, though, to using a
``quasi-random'' sequence, such as that of \citet{sobol67}; for a discussion
see \citet[][pp.~404-409]{press07}.  As the samples are accumulated, the
computation of $\langle|\nabla\phi|\rangle_{V(\psi,\Delta \phi)}$ is
straightforward.  If
$\phi(\vec{x})$ were available only at a limited number of locations then the
optimal $\Delta \phi$, based on statistical considerations, might be
determined by the method of \citet{freedman81}.  Here, however,
$\phi(\vec{x})$ is known for all $\vec{x}$ so that $\Delta \phi$ can be
selected and then $\phi(\vec{x})$ sampled until the desired statistical
convergence is attained.

In figure \ref{fig:simpleAreaConvergence} are shown the results of sampling
$\phi$ many times and accumulating the statistics for various values of
$\Delta\phi$. It is observed that the relative area in the area estimate for
a given value of $\Delta\phi$ converges with $n^{-1/2}$ where $n$ is the
number of samples.  This is expected since a pseudo-random sequence is used
for the sampling in this test case \citep[][pp.~404-409]{press07}.  The
sampling of the DNS cases, discussed in \S\ref{sec:numerical}, is a hybrid of
pseudo-random and quasi-random with somewhat faster convergence.  For each
$\Delta\phi$ a minimum relative error is reached given enough samples.
\begin{figure}
	\centering
	\includegraphics[width=\textwidth]{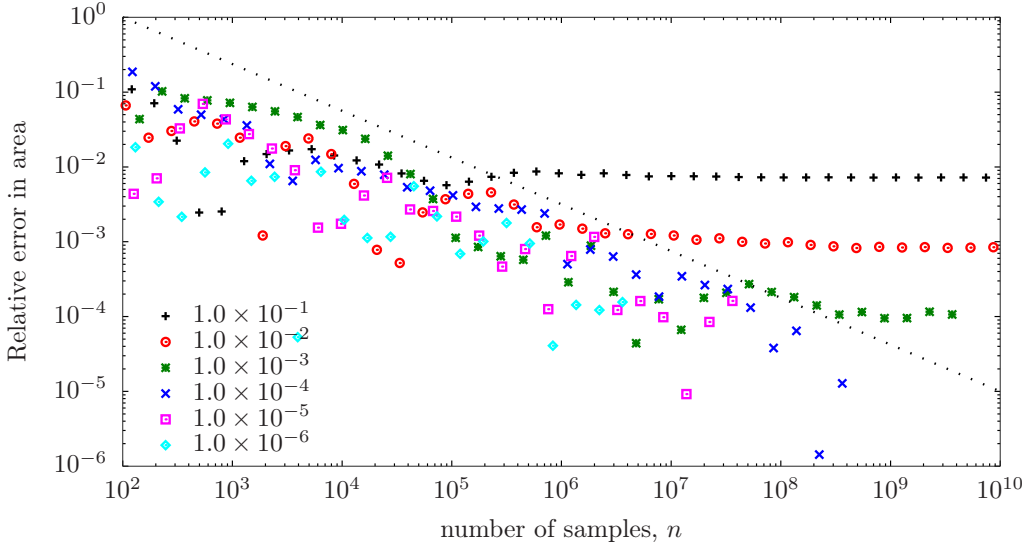}
	\caption{Relative error in the estimate for the area of an isosurface in the
		simple example with $\psi=4$.}
	\label{fig:simpleAreaConvergence}
\end{figure}
The term ``relative error'' is not strictly correct because the method arrives
at the correct areas based on layers of finite thickness.  For this simple
case, the area of the surface with $\phi = \psi + \Delta \phi$ will always be
larger than the area of the isosurface $\phi = \psi$.  For example, the
relative difference between the analytically determined areas of the surfaces
with $\psi=4$ and $4.01$ is approximately $0.001$, which is consistent with
figure \ref{fig:simpleAreaConvergence}.  Importantly, though, the area
estimate converges linearly with $\Delta \phi$ as expected from the derivation
of \eqref{eq:coarea_arearatio}.  This is apparent from the figure because, for
the three layer thicknesses for which the evaluation of
\eqref{eq:coarea_arearatio} is converged, the area estimate improves by an
order of magnitude as $\Delta \phi$ is decreased by an order of magnitude.

We conclude from this simple case that the method can be used to measure
isosurface areas with arbitrary accuracy given a sufficiently small $\Delta
\phi$ and sufficient samples.  The area converges with $n^{-1/2}$ and with
$\Delta \phi$. One caveat is that care must be used in accumulating the
statistics, such as by using 128-bit floating point arithmetic and, perhaps,
the method of \citet{kahan65} to reduce roundoff errors; both techniques are
used in this paper.
A second caveat is
that pseudo-random sequences have limitations which are beyond the scope here
but are discussed in, e.g., \citet{vadhan12}.   Also, it is interesting to note that
the quantity ${V(\psi,\Delta \phi)}/(V\ \Delta \phi)$ is in fact a first
order approximation to $P(\phi;\phi=\psi)$ provided that the Monte Carlo
integration is converged. Further discussion of this is in 
appendix \ref{sec:error_analysis_arearatio}.

\section{Direct Numerical Simulations}
\label{sec:dns}
The simulated flows are the solution to the Navier-Stokes equations and
passive scalar transport equation in three
spatial dimensions:
\begin{subequations}
	\label{eq:governeq}
	\begin{equation}
	\nabla \cdot \vec{u} = 0
	\end{equation}
	\begin{equation}
	{{\partial \vec{u}} \over {\partial t}} + \vec{u} \cdot \nabla \vec{u} = - \nabla p 
	+ \nu \nabla^2\vec{u}  + \vec{b} 
	\end{equation}
	\begin{equation}
	\frac{\partial \phi}{\partial t} + \vec{u}\cdot\nabla\phi = -
	\vec{u}\cdot\nabla\Phi + D \nabla^2 \phi \ .
	\end{equation}
\end{subequations}
The velocity vector is $\vec{u} = [u, v, w]$ with kinematic viscosity $\nu$,
the pressure $p$ has been divided through by the (constant) density, and
$\vec{b}$ is a time-varying force applied to maintain the flow statistically
stationary.  The passive scalar is decomposed into $\Phi + \phi$, with
time-invariant $\Phi$ having a uniform gradient in the direction of $w$, and
being uniform in the other spatial directions; $\phi$ denotes the fluctuations
relative to $\Phi$ with molecular diffusivity $D$.  

Spatial discretisation is via Fourier series truncated according to the 2/3
rule to remove all effects of aliasing.  The advection terms are computed in
real space in vorticity form for momentum and in convective and conservation
forms on alternating time steps for the scalars.  The equations are advanced in
time using a third order accurate fractional step method for the advection 
and pressure gradient terms while the diffusion terms are integrated exactly in
Fourier space.

The velocity fields are forced isotropically so that the three-dimensional
kinetic energy spectrum at length scales larger than the integral length scale
matches Pope's model spectrum with his $p_0=2$ and his $c_L=6.78$
\cite[(equation 6.247)]{pope00}.  Each time step, a second-order ordinary
differential equation is advanced in time for each shell of the shell-averaged
spectrum to determine how much energy to add to that shell.  The energy is
distributed randomly, but consistent with continuity, across all the wave
numbers composing the shell.  The solution to these differential equations and
the random distributions produces the Fourier-space equivalent of $\vec{b}$.
This approach is introduced by \citet{overholt98} to efficiently converge a
simulation to a target spectrum; \citet{rao11} reviews a variety of forcing
techniques for the type of simulation used for this research.
After the simulation was determined to be statistically stationary based on
mean quantities, it was advanced for an additional large eddy turnover time
before taking a snapshot for analysis of the isosurface areas. 

Seven velocity fields are considered for this research, with each convecting
four scalar fields with differing Schmidt numbers $\Sc=\nu/D$.
The velocity field statistics are tabulated in table \ref{tbl:tableOne} and the scalar
statistics in table \ref{tbl:tableTwo} 
\begin{table}
	\begin{center}
		\begin{tabular}{l r r r r r r r}
 & \multicolumn{1}{c}{R24} &   \multicolumn{1}{c}{R42} &   \multicolumn{1}{c}{R98} &   \multicolumn{1}{c}{R154} &   \multicolumn{1}{c}{R245} &   \multicolumn{1}{c}{R400} &   \multicolumn{1}{c}{R633} \\\midrule $\ReL$ &49.8 & 114 & 310 & 734 & 1853 & 4790 & 12150 \\ $\Rey$ &24.0 & 42.3 & 97.8 & 154 & 245 & 400 & 633 \\ ${\cal L}/L$ &5.5 & 6.7 & 5.2 & 5.7 & 5.7 & 5.6 & 5.6 \\ $\kappa_{max} L_k$ &9.73 & 9.14 & 26.84 & 12.90 & 6.48 & 9.39 & 5.65 \\ $L_k / \Delta$ &4.66 & 4.38 & 12.82 & 6.16 & 6.64 & 4.48 & 2.70 \\ $\tau_k / \Delta t$ &  90 &  103 & 1376 & 1165 &  954 & 1170 & 1202 \\ $N$ &512 & 1024 & 4096 & 4096 & 8800 & 11760 & 14256\\ Reference & & & R1 & R2 & R3 & R4\\\end{tabular}
	\end{center}
	\caption{Velocity parameters and statistics for the highest resolution of
		each case. $N$ and $\kappa_{max}L_k$ are shown for each velocity field at it
		highest resolution ($\Sc=7$); lower resolution is used
		for the cases with lower $\Sc$ in simulations R245, R400, and
		R633 and the resolution for each scalar case is given in table
		\ref{tbl:tableTwo} in terms of $\Lk/\Delta$. `Reference' indicates the names used for statistically
		equivalent simulations in \citet{almalkie12}, \citet{debk15}, \citet{muschinski15}
		and \citet{portwood16}.
		\label{tbl:tableOne}}
\end{table}
\begin{table}
	\begin{center}
		\begin{tabular}{r r r r r r r r}
$Pe_\lambda$ & $Sc$ & $L_k/\Delta x$ & $L_c/\Delta x$ & $L_b/\Delta x$ & $\tau_k/\Delta t$ & $\tau_c/\Delta t$ & $Re_\lambda$ \\\midrule2.40 & 0.10 & 4.66 & 26.23 & 14.75 & 89.95 & 284.43 &  24\\
4.23 & 0.10 & 4.38 & 24.63 & 13.85 & 103.09 & 326.01 &  42\\
9.78 & 0.10 & 12.82 & 72.08 & 40.53 & 1376.14 & 4351.74 &  98\\
15.37 & 0.10 & 6.16 & 34.64 & 19.48 & 1164.67 & 3683.00 & 154\\
\\
16.82 & 0.70 & 4.66 & 6.09 & 5.57 & 89.95 & 107.51 &  24\\
24.03 & 1.00 & 4.66 & 4.66 & 4.66 & 89.95 & 89.95 &  24\\
24.46 & 0.10 & 4.53 & 25.48 & 14.33 & 953.77 & 3016.08 & 245\\
29.58 & 0.70 & 4.38 & 5.72 & 5.23 & 103.09 & 123.22 &  42\\
\\
40.01 & 0.10 & 2.22 & 12.51 & 7.03 & 485.56 & 1535.47 & 400\\
42.25 & 1.00 & 4.38 & 4.38 & 4.38 & 103.09 & 103.09 &  42\\
63.27 & 0.10 & 1.55 & 8.71 & 4.90 & 779.68 & 2465.55 & 633\\
68.47 & 0.70 & 12.82 & 16.75 & 15.32 & 1376.14 & 1644.80 &  98\\
\\
97.81 & 1.00 & 12.82 & 12.82 & 12.82 & 1376.14 & 1376.14 &  98\\
107.59 & 0.70 & 6.16 & 8.05 & 7.36 & 1164.67 & 1392.04 & 154\\
153.70 & 1.00 & 6.16 & 6.16 & 6.16 & 1164.67 & 1164.67 & 154\\
168.24 & 7.00 & 4.66 & 1.08 & 1.76 & 89.95 & 34.00 &  24\\
\\
171.22 & 0.70 & 4.53 & 5.92 & 5.41 & 953.77 & 1139.97 & 245\\
244.60 & 1.00 & 4.53 & 4.53 & 4.53 & 953.77 & 953.77 & 245\\
280.05 & 0.70 & 2.22 & 2.91 & 2.66 & 485.56 & 580.35 & 400\\
295.76 & 7.00 & 4.38 & 1.02 & 1.66 & 103.09 & 38.97 &  42\\
\\
400.07 & 1.00 & 2.22 & 2.22 & 2.22 & 485.56 & 485.56 & 400\\
442.86 & 0.70 & 1.55 & 2.02 & 1.85 & 779.68 & 931.89 & 633\\
632.65 & 1.00 & 1.55 & 1.55 & 1.55 & 779.68 & 779.68 & 633\\
684.65 & 7.00 & 12.82 & 2.98 & 4.84 & 1376.14 & 520.13 &  98\\
\\
1075.87 & 7.00 & 6.16 & 1.43 & 2.33 & 1164.67 & 440.20 & 154\\
1712.20 & 7.00 & 6.64 & 1.54 & 2.51 & 953.77 & 360.49 & 245\\
2800.48 & 7.00 & 4.48 & 1.04 & 1.70 & 1170.48 & 442.40 & 400\\
4428.57 & 7.00 & 2.70 & 0.63 & 1.02 & 1202.44 & 454.48 & 633\\
\\
\end{tabular}
	\end{center}
	\caption{Scalar field parameters and statistics all cases.
		\label{tbl:tableTwo}}
\end{table}
with $\ReL$ the Reynolds number based on the
r.m.s.\ velocity and the integral length scale $L$, $\Rey$ the Reynolds number
based on the r.m.s.\ velocity and the Taylor microscale and $\Pey$ the
corresponding P\'eclet number.  $\Lk$, $\Lb$, and $\Lc$ are the Kolmogorov,
Batchelor, and Oboukhov-Corrsin length scales, respectively,
\begin{equation}
L_k=\left(\frac{\nu^3}{\epsilon}\right)^{1/4} , \ \ \
L_b=\left(\frac{\nu D^2}{\epsilon}\right)^{1/4}, \ \ \
L_c=\left(\frac{D^3}{\epsilon}\right)^{1/4}
\end{equation}
with corresponding times scales
\begin{equation}
\tau_k=\frac{\nu}{\epsilon}^{1/2} , \ \ \
\tau_c=\frac{D}{\epsilon}^{1/2} \ .
\end{equation}
Here $\epsilon$ is the dissipation rate of kinetic energy.
The maximum wave number after dealiasing is $\kappa_{max}$, the grid spacing
is $\Delta x$, and the time step size is $\Delta t$.  The domain size is $\cal
L$ in each spatial direction discretised by 
$N$ grid points.

The velocity fields are denoted by an R followed by $\Rey$ rounded to an
integer, e.g. R24.  Four of the cases have been reported elsewhere so that
cases R98, R154, and R245 are statistically equivalent to cases R1, R2, and R3
reported by \citet{almalkie12} and R400 is equivalent to the homogeneous
isotropic case denoted R4 in \citet{almalkie12a},
\citet{debk15} and \citet{portwood16}. The R400 is denoted R4 by
\citet{muschinski15} and used to study models for scalar energy spectra.

\section{Numerical considerations}
\label{sec:numerical}
Before proceeding to evaluating isosurface areas in the DNS as a function
of Reynolds and Schmidt numbers in \S\ref{sec:results}, it is important to
apply the tests from \S\ref{sec:simple} to understand the accuracy to which
the areas can be evaluated from the DNS.  It is also important to understand
the extent to which the spatial and temporal resolution of the DNS may effect
the area calculations.  These are the topics of this section.  We do not
review the numerical errors in Fourier spectral methods because they are
discussed so extensively in the literature and instead refer the reader to,
e.g., \citet[][\S 3]{debk01c}.

Fourier spectral simulations represent the scalar fields as infinite series
with all but a few of the series coefficients identically zero.  This is in
contrast to, say, a laboratory time series which has been Fourier transformed
and represented by a truncated series.  This characteristic of the current
simulations allows us to sample the scalar fields at an arbitrary number of locations
via phase shifting so that each sample has the same accuracy as the simulation
as a whole and no interpolation error is introduced.  Also in Fourier spectral
simulations, $\nabla \phi$ is known with the same accuracy as $\phi$ because
the derivative of each term in the series expansion is known analytically.  

The sampling process begins with the scalar field in Fourier space.  A random
phase shift with wave number vector $\vec{\kappa}$ is applied, the field is
transformed to real space, and all of the points in the resulting field are
used to accumulate data for evaluating \eqref{eq:coarea_arearatio}.  The process is
repeated as many times as desired.  The result is a Monte Carlo method in
which for every $N^3$ samples of the field only one random vector $\kappa$ is
chosen.  Because the samples for a single value $\vec{\kappa}$ are uniformly
distributed across the entire simulation domain, this method is a type of
``quasi-random'' sampling (c.f.~\S\ref{sec:simple}).

The spatial and temporal resolution requirements in DNS have been continually
refined over the last three decades as computers have gotten more powerful so
as to enable simulations with higher internal intermittency.  The large-scale
requirement for decaying homogeneous isotropic turbulence (HIT) was established by
experimentation to be about 20 times the integral length scale $L$ by
\citet{debk98} and recently verified by \citet{debk19}.  In simulations forced
to be statistically stationary, such as those used for the current research,
no large-scale resolution requirement has been established to our knowledge,
and we have used domains that are approximately $5 L$ because this
resolution has been shown to produce structure functions and spectra that
match theoretical predictions in the inertial and dissipation ranges
\citep{almalkie12}.

Small scale resolution requirements for DNS were first established by
\citet{eswaran88}, who showed that $\kappa_{max} \Lk > 1$ is required for the
scalar variance to decay as expected from theory in decaying HIT.  Since then,
the dependency of small scale statistics on Reynolds number has been analysed
in several studies \citep{yeung02a,yeung05,ishihara07,gulitski07c,
  gulitski07b,ishihara09} and the effects on intermittency of insufficient
small-scale resolution in DNS
investigated~\citep{yakhot05,schumacher07,watanabe07, donzis08}.  Recently,
\citet{yeung18} report on not only the spatial resolution required to resolve
small-scale intermittency but also the temporal resolution for $\Rey=390$ and
650, which are comparable to our cases R400 and R633.  Because of our need to
resolve $\Sc=7$, our spatial resolutions are finer than those shown to be
sufficient in the paper just referenced.  Therefore, when we satisfy the
Courant number condition established by Yeung et al.\ our timestep is smaller
than theirs, which gives us confidence that the time scale of intermittency is
resolved.  The time step in each simulation is given in table
\ref{tbl:tableOne}, relative to the Corrsin and Kolmogorov time scales, from
which it is apparent that these time scales are very well resolved in the
simulations.

The foregoing gives us confidence that all the scalar and velocity fields are
well resolved and so we move to the question of the effect of layer thickness
$\Delta \phi$ used in numerically evaluating \eqref{eq:coarea_arearatio}.  For
the simple example illustrated by figure \ref{fig:simpleAreaConvergence}, the
error in an isosurface area reduces unbounded and proportionally to
$n^{-1/2}$ and $\Delta \phi$. Recall that the exact area in that case is
known.  In the DNS fields, though, the exact area is not known so that the
error in the area must be based on the best available area estimate, e.g.,
that computed with the scalar field at the highest resolution available and
with the smallest practicable layer thickness.  Also, we can surmise that, even with
careful attention to mitigating them with 128-bit arithmetic and the method of
\citet{kahan65}, roundoff errors will limit the accuracy of the numerical
evaluation of \eqref{eq:coarea_arearatio}.

Our test case for numerical studies is R245 because the Reynolds number is high
enough for there to be significant intermittency yet we have resolved it with
much finer spatial resolution than required per \citet{yeung18}.  $\Pr=7$ is
considered since it has the smallest diffusive length scale.  The relative
errors for the area of a single isosurface $\psi$ computed using various layer
thicknesses is shown as figure \ref{fig:layerthickness}.  The dimensionless
layer thickness is
\begin{equation}
\widetilde{\Delta \phi} = \frac{\Delta \phi}{\Lr \left<|\nabla\phi|\right>_{\Xpsi}} 
\end{equation}
with $\Lr$ being some resolution length scale appropriate for the
scalar.  In this paper we set $\Lr=\Delta x$ because, in our simulations,
$\Delta x < \Lb$ for $Pr \ge 1$ and $\Delta x \le \Lc$ for $Pr \le 1$.

\begin{figure}
	\centering
	\includegraphics*{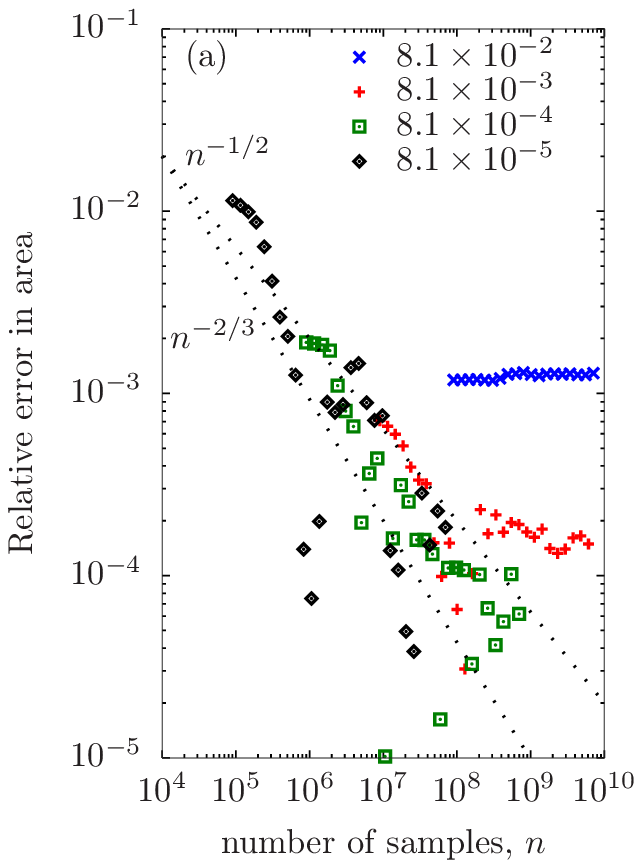}
	\includegraphics*{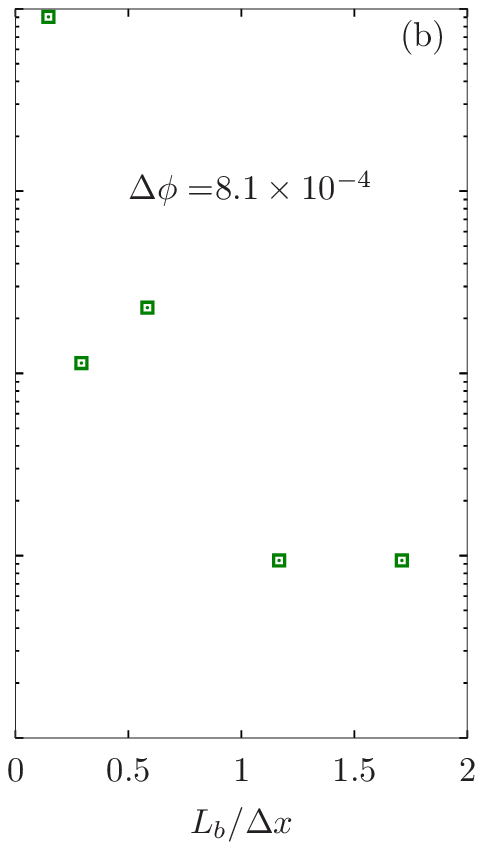}
	\caption{Error in the area of one isosurface in case R245, $Pr=7$, for
		various nondimensional layer thicknesses $\widetilde{\Delta\phi}$ (panel
		(a)) and grid resolutions (panel (b)).  The dotted lines
                labelled $n^{-1/2}$ and $n^{-2/3}$ mark the convergence rates
                expected for pseudo-random and quasi-random sampling,
                respectively \citep{press07}.
		\label{fig:layerthickness}}
\end{figure}

The first thing to note from the figure \ref{fig:layerthickness}(a) is that the
leftmost symbol in each set is the number of samples $n$ in a layer of
thickness $\widetilde{\Delta \phi}$ before phase shifting, e.g., when
$\widehat{\Delta \phi} = 8.1\times10^{-5}$ approximately $10^5$ points are in
$[\phi, \psi+\Delta\phi)$.  Next it is noted that when $\widetilde{\Delta \phi} =
8.1\times10^{-2}$, the minimum relative error is approximately $10^{-3}$ and
does not decrease with more sampling.  For smaller $\widetilde{\Delta \phi}$,
the isosurface area estimates converge at a rate between $n^{-1/2}$, which
is expected for pseudo-random sequences, and $n^{-2/3}$, which is expected for
quasi-random sequences.  Our sampling method is based on phase shifting the
entire field based on a randomly selected $\kappa$ so that the sampling for a
single phase shift is quasi-random but the phase shifts are pseudo-random.

Given enough samples for a given $\Delta \phi$ the area estimate converges
just as it does in \S\ref{sec:simple}. For example, when $\widetilde{\Delta
  \phi} = 8.1\times10^{-3}$ then the minimum relative error is approximately
$1\times10^{-4}$ meaning that the isosurface area estimate converges to a
value different from the ``exact'' value taken to be the estimate when
$\widetilde{\Delta \phi} = 8.1\times10^{-5}$.  As explained in
\S\ref{sec:simple}, the term ``relative error'' is not quite correct because
the method computes the area based on a layer of finite thickness so that the
area based on $\widetilde{\Delta \phi} = 8.1\times10^{-3}$ is expected to
be different from that based on $\widetilde{\Delta \phi} = 8.1\times10^{-5}$
even if both areas are computed exactly.

From figure \ref{fig:layerthickness}(a) we conclude that the error in the area
estimate can be made arbitrarily small given a small enough layer thickness
and a sufficient number of samples produced by phase shifting.  This
conclusion is conditioned on the assumption that the simulations are
sufficiently resolved at the small scale.  We have already concluded that we
expect them to be, based on the requirements determined by \citet{yeung18},
but we verify it one more time with figure \ref{fig:layerthickness}(b).  The
``exact'' area for this plot is from a simulation with $\Lb/\Delta x = 2.5$
that is filtered with a spectral cutoff filter to produce the fields with
lower values of $\Lb/\Delta$ used to make the figure.  It is observed that the
error in the area estimate converges to its minimum value when $\Lb/\Delta x =
1$.  Note from figure \ref{fig:layerthickness}(a) that this minimum value of
approximately $1\times 10^{-4}$ is set by the layer thickness, not the
resolution of the simulation.

\section{Effects of $\bm{\Rey}$ and $\bm{Sc}$ on isosurfaces}
\label{sec:results}
The time rate of change of the area of a non-material isosurface is
determined by the balance between stretching tangential to the surface and
diffusion normal to the surface.  A detailed derivation is included in
\citet{dopazo18} for the case of a chemically reacting flow.  For our purpose
here we follow their notation and consider an infinitesimal piece of an
isosurface having area $\cal S$ with normal vector $n_i$ and strain rate
$S_{ij}$ so that the strain rate tangential to the surface is $a_T =
(\delta_{ij} - n_i n_j) S_{ij}$.  A point on an isosurface moves by
diffusion, relative to the speed at which the surface is advected, with speed
$S_d = - D\nabla^2\phi / \nabla \phi$.  Therefore the fractional change in
area of this piece of an isosurface is \citep[c.f.][equation 25]{dopazo18}
\begin{equation}
\frac{1}{\cal S} \frac{\partial \cal S}{\partial t} = a_T + S_d \frac{\partial
	n_i}{\partial x_i} \ .
\label{eq:dSdt}
\end{equation}
For the statistically stationary flows, the integral of \eqref{eq:dSdt} over
an entire isosurface is zero meaning that the integrals of two terms on the
right hand side of \eqref{eq:dSdt} come into equilibrium.

The central physics question motivating this paper is how the equilibrium
surface area $A_\psi$ is affected by Reynolds and Schmidt
number. \citet{catrakis02} measured area-volume properties of fluid interfaces
in a laboratory water jet with jet Reynolds number $\sim 20,000$ and $Sc \sim
2000$ and report that, while large and small length scales contribute to the
area-volume ratio, the dominant effect is at the small scales.  They
conclude that self-similarity at the small scales can be expected to enable
extrapolation of area-volume behaviour to higher Reynolds number.
\citet{schumacher05a} consider DNS of a passive scalar with a mean gradient in
homogeneous isotropic turbulence with $10 \le \Rey \le 42$ and $2 \le Sc \le
32$ and observe that the area-volume ratio increases with both $Sc$ and $\Rey$
but that a power law is not observed. They further note that power law
behaviour is not expected because, at the Reynolds and Schmidt numbers
considered, neither the inertial-convective nor the viscous-convective
subranges
exist.  \citet{oneill05} also consider DNS of homogeneous isotropic turbulence
with a mean scalar gradient with $Sc=0.25$ and 5.0 and $\Rey=41$, 42, and 77.
They observe that increasing either $\Rey$ or $Sc$ results in the size of the
isoscalar surfaces decreasing.  We note that they determine the areas with a
geometric method and review in \S1 some of the challenges with using geometric
methods to find the areas of turbulent isosurfaces.

In the literature just cited, the metric for isosurface area is the
area-volume ratio.  Our simulations are computed in periodic domains, however,
so that the scalar can be convected out of the domain to a periodic domain
above or below.  Given this configuration, we find it to be more clear to take
advantage of homogeneity in the flows and consider the dimensionless ratio of
the wrinkled surface to the initial surface area, which is that of a plane
normal to the mean scalar gradient.  The wrinkled surface area includes that
of any portions of the surface advected across the periodic boundaries.  To be
precise, we compute the volume of a very thin wrinkled layer with
$\widetilde{\Delta \phi} = 1 \times 10^{-4}$ and relate it to an isosurface
area via \eqref{eq:coarea_arearatio}.  The results are in figure
\ref{fig:peclet}.
\begin{figure}
	\centering
    \includegraphics{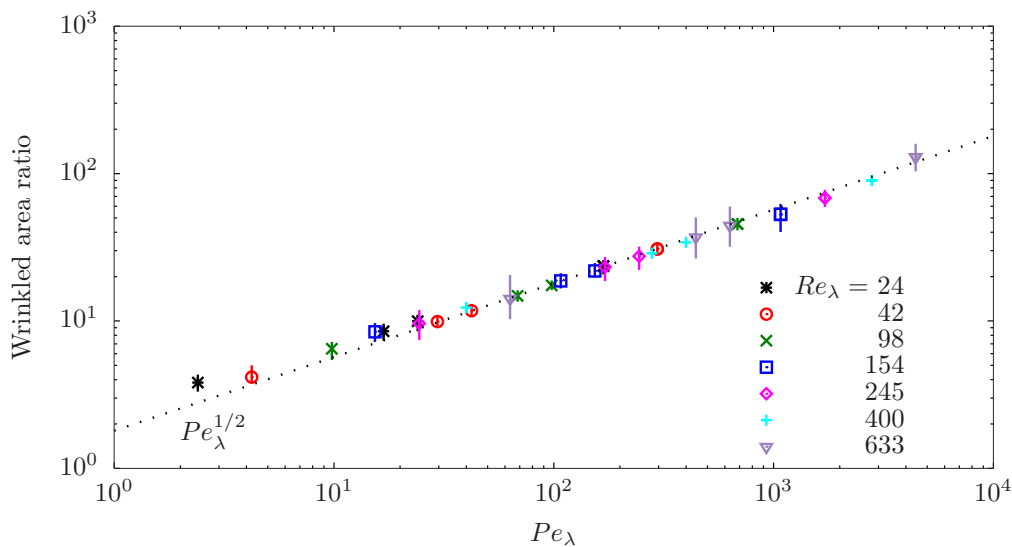}
	\caption{
		Isosurface areas normalised by the area of a horizontal plane
		in the simulations.  The symbols mark the average for 20 isovalues
		and the bars indicate the range from the largest to smallest areas.
		The dotted line is a reference line, not a fitted line.
		\label{fig:peclet}
	}
\end{figure}
For each case, the areas are measured for 20 iso-values spanning the range of
$\phi$ in the simulation.  It is noted that there is more scatter in some
cases than in others, and this appears to be due to limited large-scale
resolution so that large structures occur which are resolved in space but not
resolved statistically.  In appendix B, we demonstrate for several simulation
cases that the scatter in the areas is much smaller simulation domain
is eight times larger.

The provisional conclusion drawn from figure \ref{fig:peclet} is that the area
increases with $Pe_\lambda^{1/2}$ for $Pe_\lambda^{1/2} > 50$ and may deviate
from power law scaling for $Pe_\lambda^{1/2} < 50$. The second part of this
conclusion is influenced by the finding by \citet{schumacher05a} that the area
increases with P\'eclet number but with no power law scaling; they consider
flows with lower 
$\Rey$ than all but our cases R24 and R48.  We consider this conclusion in
more detail in the remainder of this section, but we immediately note that it
is not in agreement with the results of \citet{oneill05} who report a decrease
in area with increasing Reynolds and Schmidt number. Their result, however, is
based on visual geometric methods, the limitations of which we discuss in
\S1. Our results are in broad agreement with the conclusion of
\citet{catrakis02} that small-scale similarity is expected; note that the
current flow configuration does not allow us to compare with their results
regarding variations in large-scale features of the flow, but we do consider
the effects of large scale resolution in appendix B and show that the average
isosurface area does not change with domain size but the scatter in the areas
of different isosurfaces in a simulation decreases as the domain size
increases and the large-scale statistics improve.

Our range of $\Pey$ overlaps that of \citet{schumacher05a}, but,
except for our cases R24 and R42, our Reynolds numbers are much higher.  So
now let us consider our conclusions from the previous paragraph in more detail
and, in particular, the point made by \citet{schumacher05a} that they do not expect to see
fractal-like behaviour of an isosurfaces because of the absence of
inertial-convective and the viscous-convective subranges.  Definitive
identification of these subranges in a simulation is difficult because of
noisy statistics in the spectra and structure functions and the presence of
the so-called spectral blocking ``bump'' at the top of the dissipation range;
\citet{ishihara09,ishihara16} provide reviews.  We can say with confidence
that we do not expect an inertial subrange in our cases R24, R42, and R98 but
that subranges do exist in cases R245, R400, and R633.  This is based not only
on detailed analysis of our data \citep{almalkie12,debk15} but also of
simulations by \citet{yeung05} and the two papers just cited by Ishihara et
al.  Yeung et al.\ show very clear inertial subranges for $\Rey \ge 240$,
which corresponds to our cases R245, R400, and R633.  Given the robust
literature on the existence of inertial ranges as a function of $\Rey$, we do
not show the inertial range analyses for all of our data sets here.  We do so next
for our case with the highest P\'eclet number because it has not been previously
published.

The compensated shell-averaged spectra of kinetic and scalar energies for the
case with $\Pey=4429$ are shown in figure \ref{fig:spectra}
\begin{figure}
	\centering
    \includegraphics*{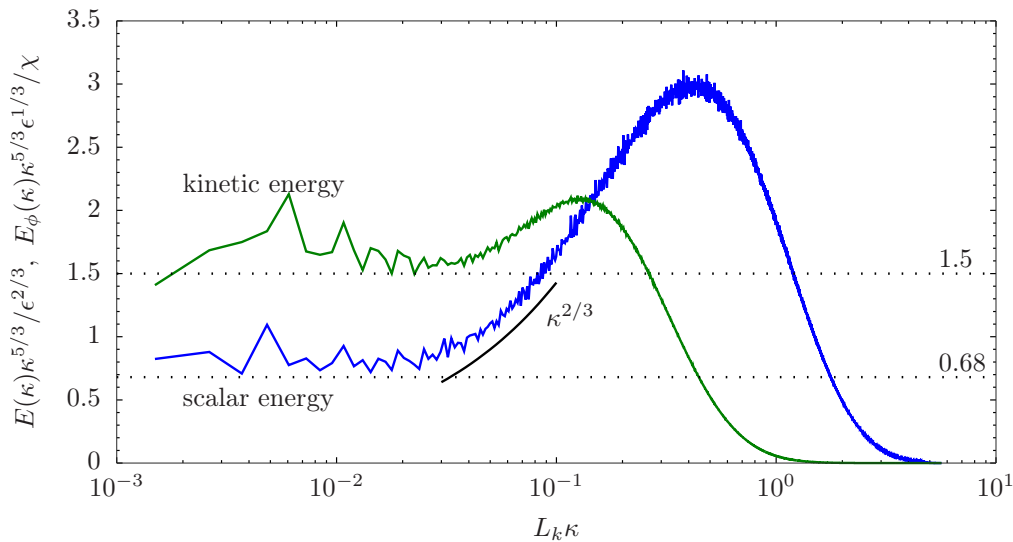}
	\caption{ Spherical shell-averaged spectra of kinetic and scalar
          energies for the case with $\Pey=4429$.  The dotted lines indicated
          the expected levels for the inertial-convective range discussed in
          the text.  The solid like marked $\kappa^{2/3}$ is the 'slope' of
          the compensated Batchelor spectrum, not a fitted curve.
		\label{fig:spectra}
	}
\end{figure}
with $E(\kappa)$ and $E_\phi(\kappa)$ the kinetic and scalar energy spectra,
respectively, $\epsilon$ the domain-averaged dissipation rate
of kinetic energy, and $\chi$ the corresponding value for scalar variance.
The textbook value for the Kolmogorov constant of $C=1.5$ \citep{pope00} is
indicated by a horizontal dotted line on the figure.  The Oboukhov-Corrsin
constant $C_\phi=0.68$ is determined by \citet{muschinski15} for a simulation
very similar to the current cases with $\Pey=280$ and 400 and 
is consistent with values determined by \citet{sreenivasan96} and
\citet{watanabe04}; it is also marked by a horizontal dotted line on the
figure.  We conclude that the inertial-convective subrange spans about a
decade in wave number based on the range over which the plateaus in both
spectra corresponds with the appropriate dotted line.  The span of of the
subrange is consistent with the results of \citet{yeung05} for their case with
$\Rey=700$.  

The viscous-convective subrange is not as easy to identify because of the
so-called ``Hill bump'' that includes this subrange and continues into the
dissipation range \citep{hill78}. The viscous-convective range, if it exists,
is expected to begin at about $L_k \kappa =0.03$ \citep{muschinski15}.  If the
viscous-convective subrange exists then we expect to see the compensated
scalar spectrum scaling with $\kappa^{2/3}$ and this scaling is marked on the
figure with a solid line labelled accordingly.  It appears that the scalar
spectrum exhibits this scaling over a fraction of a decade of wave numbers in
the viscous-convective subrange.  Of course the spectrum necessarily
transitions from the plateau to the Hill bump in the dissipation range and so
it is difficult to clearly identify a narrow viscous-convective subrange,
which is why we do not present this analysis for the cases with lower $\Pey$.

Next, let us consider the probability density functions of the local
scalar gradient on an isosurface plotted as figure \ref{fig:grad} for five
layers in each case.  
\begin{figure}
	\centering
    \includegraphics{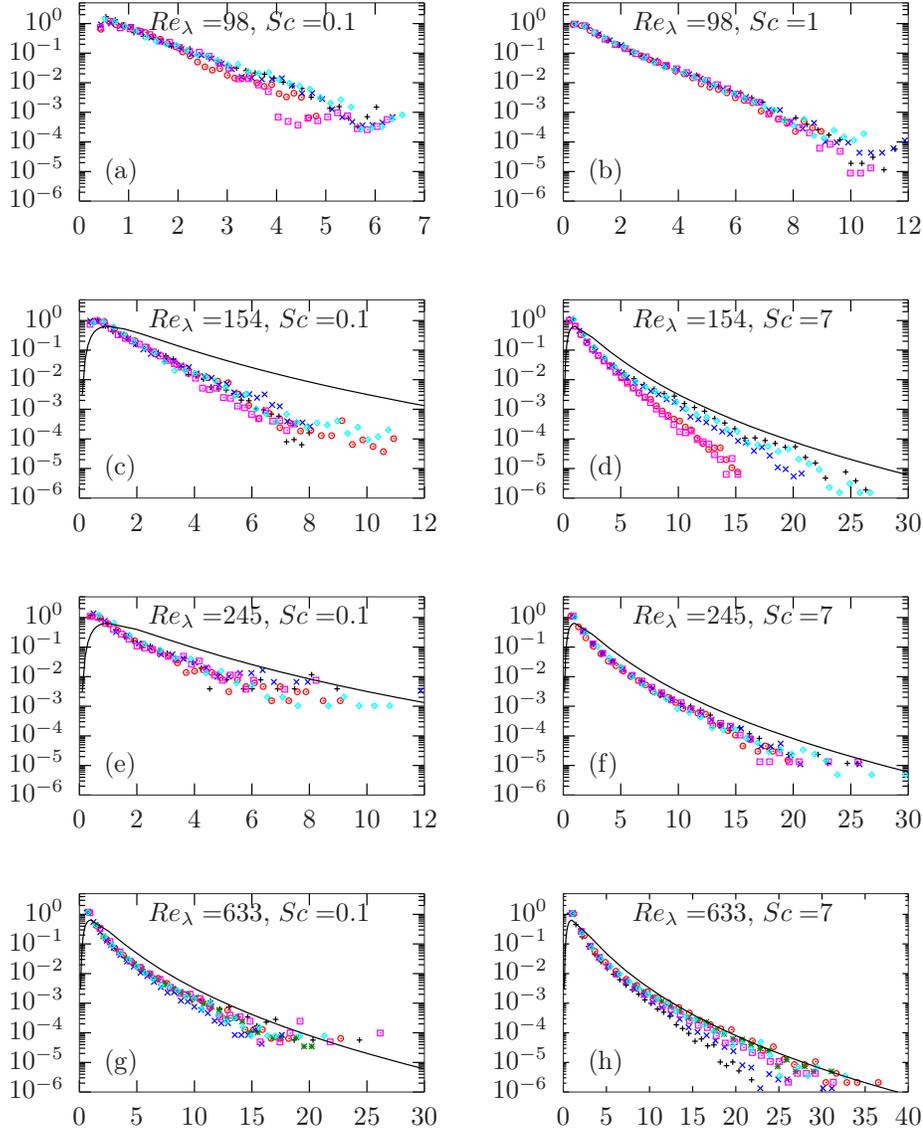}
	\caption{ Probability density of the local gradient magnitudes on a layer
          normalised by the mean for that layer.
          The horizontal axis is
          $|\nabla\phi|_{V(\psi,\Delta
            \phi)} /
          \langle|\nabla\phi|\rangle_{V(\psi,\Delta
            \phi)}$.  Five layers are shown for each case with their nominal
          locations distributed uniformly over the vertical extent of the
          domain in this order of lowest to highest $\phi$: plus, circle,
          cross, square, diamond.  In panels (c) though (h), the solid line marks
          the standard log-normal distribution.
		\label{fig:grad}
	}
\end{figure}
In panels (a) and (b) the distributions are very close to exponential,
consistent with the results of \citet{schumacher05a}.  In panels (e), (f),
(g), and (h) they tend toward lognormal with the lognormal distribution being
a very good model for some of the layers in panel (h).  In panels (c) and (d),
it is difficult to draw a conclusion although in our opinion the curves are
neither straight enough to be consider exponential nor is the lognormal
distribution a good model.  Note that the change in the distribution from
exponential to one closer to lognormal is appears primarily to be due to
increased $\Rey$ largely independent of either $\Pey$ or $Sc$.  This suggests
that the exponential distribution occurs when there is no inertial-convective
subrange.  Recall that case R98 has no inertial subrange whereas our cases
R245, R400, and R633 have characteristics of inertial subranges.

\begin{figure}
  \centering
  \includegraphics{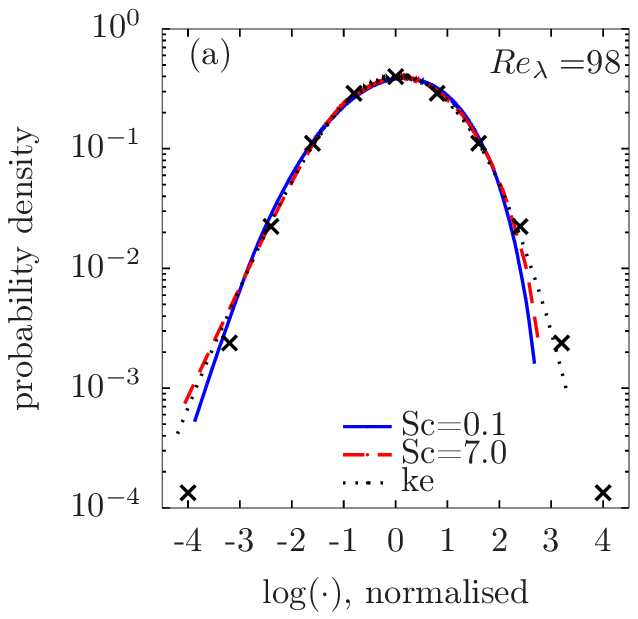}
  \hspace*{-0.3in}
  \includegraphics{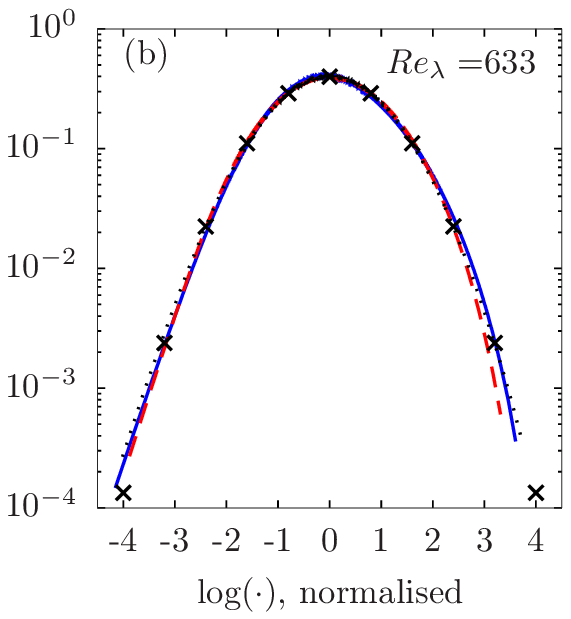}
  \caption{\label{fig:chi} P.d.f.s of the logarithm of the dissipation rates
    for the variances of scalars with $Sc=0.1$ and 7.0 and for kinetic
    energy, all normalised by their means and variances.  The crosses mark the standard Gaussian.}
\end{figure}

Significant scatter is observed figure \ref{fig:grad} among the distributions
for the five layers in each set.  The governing equations we solve numerically
describe a homogeneous flow but in a simulation domain of finite size the
homogeneity is not perfect.  In appendix B we conclude that the areas of
different isosurfaces converge to a common value as the domain size increases.
So the effects of limited domain size observed in the scatter in figure
\ref{fig:grad} are interesting in itself.  It is most clear in panel (d) in
which the lowest curve is approximately exponential while the upper curve is
approximately lognormal.  Recall from the literature reviewed in this section
that an inertial-convective range is expected for $\Rey>240$ but not
necessarily at $\Rey=154$, which is the value for panels (c) and (d).  The
result is consistent with $\Rey>154$ being between $\Rey=98$, for which an
inertial-convective range is not expected, and $\Rey=245$ for which one is
expected.

A different view of the same data is presented in figure \ref{fig:chi}.  More
precisely, the p.d.f.s are shown for the natural logarithm of the local
dissipation rate of scalar variance $\chi_0 = 2D \nabla \phi \cdot \nabla
\phi$ using data for the whole domain.  By plotting this quantity, comparisons
can be made with the logarithm of the p.d.f.\ of the local dissipation rate of
kinetic energy $\epsilon_0$.  For $\Rey=98$, $\epsilon_0$ has an approximately
lognormal distribution, except for very small values of $\epsilon_0$, whereas
the scalar does not.  For both cases shown with $\Rey=98$ and for $\Rey=633$
with $Sc=0.1$ we observe that the p.d.f.\ for high values of $\chi_0$ lies
inside that of $\epsilon_0$.  This behaviour was first noted, to our
knowledge, in laboratory measurements in boundary layers \citep{antonia77} and
later in in DNS of homogeneous isotropic turbulence \citep[e.g.][]{vedula01}.
For $\Rey=633$ and $Sc=7.0$, however, this behaviour occurs only at the very
highest values of $\chi_0$.  This suggests that, if the P\'eclet number were
high enough then the lognormal model would be a good approximation for
$\chi_0$, except for very small values at which the p.d.f.s of $\chi_0$ and
$\epsilon_0$ depart from lognormality but agree with each other.

\section{Conclusions}
The area of a scalar isosurface can be written exactly in terms of the limit
of the volume of a thin layer as the layer thickness goes to zero, which is
shown in the development of \eqref{eq:coarea_arearatio}.
Numerically evaluating \eqref{eq:coarea_arearatio} to arbitrarily small
relative error in DNS via a Monte-Carlo method requires being able to
evaluate accurately the scalar field at arbitrary locations, which can be done
in the case of pseudospectral simulations.  We demonstrate the evaluation of
the surface areas for an analytical test case and then with DNS data and
observe the expected convergence behaviours as the thickness of the layer is
reduced and the number of samples increases.  For the DNS, it is observed that
with sufficient spatial resolution of the simulation and a sufficiently thin
layer width that the relative error in the area of an isosurface can be
reduced to $1\times10^{-4}$ or smaller.  In the process of determining this,
we verify the small-scale
resolution criterion recently provided by \citet{yeung18}.

Next the methodology is applied to 28 DNS cases of statistically stationary
homogeneous isotropic turbulence with a mean scalar gradient, $0.1 \le Sc \le
7$ and $24.0 \le \Rey \le 633$.  It is observed that the layer area is very
nearly proportional to $\Pey^{1/2}$ over the range $50 \le \Pey \le
4429$ and may depart from this scaling when $\Pey < 50$.
This result is broadly consistent with the literature although direct
comparisons are not possible because other studies used geometric estimates
for isosurface areas rather than direct calculations. 

To understand the results better, we expand on analysis by \citet{schumacher05a} and
examine the p.d.f.s of the scalar gradient on multiple layers in each
simulation and observe that at low $\Rey$ they are modelled very well by an
exponential distribution.  For our highest $\Rey$ case, the lognormal
distribution is a much better model, and we also see convincing evidence of
the existence of a viscous-convective subrange in this case.  Indeed, for
$\Rey=633$ and $Sc=7.0$, the lognormal model is very good for the p.d.f.\ of
the scalar dissipation rate, and the difference between the p.d.f.\ of the
scalar and kinetic energy dissipation rates is extremely small.  These
results, combined with those of \citet{schumacher05a} suggest that the
isosurface area scales with $\Pey^{1/2}$ only if there is an
inertial-convective subrange but does not require there to be a
viscous-convective subrange.  The new results also suggest that the difference
between the p.d.f.s\ of the scalar and kinetic energy dissipation rates, which
has long been observed, might occur only when the Reynolds number is not
sufficiently high.

\label{sec:conclusions}

\section*{Acknowledgements}
We thank Professors Duane Storti and Jim Riley for their valuable insights.
High
performance computing resources were provided through the U.S.\ Department of
Defense High Performance Computing Modernization Program by the Army Engineer
Research and Development Center and the Army Research Laboratory under
Frontier Project FP-CFD-FY14-007.

\appendix
\section{Relationship between $\bm{ A(\psi)/V}$ and $\bm{ P(\phi;\phi=\psi)}$}
\label{sec:error_analysis_arearatio}
At the end of \S 3 we note that the the ratio ${V(\psi,\Delta\phi)}/(V\ \Delta \phi)$ is an approximation to $P(\phi;\phi=\psi)$.
We show here our
rationale for that statement and also present an analysis for the order of
accuracy of \eqref{eq:coarea_arearatio}. Recall from \S\ref{sec:arearatio}
that $A(\psi)$ denotes the exact area of the surface with $\phi=\psi$. The
ratio ${V(\psi,\Delta \phi)}/{V}$ is the ratio of the number of locations with
scalar value between $\psi$ and $\psi + \Delta \phi$ to the total number of
locations, which equals the probability $p(\phi;\psi \leq \phi < \psi + \Delta
\phi)$ of finding a point with scalar value in the range $[\phi,\phi + \Delta
  \phi)$. This equivalence of ${V(\psi,\Delta \phi)}/{V}$ with the probability
  of finding a position $\vec{x}$ such that $\phi(\vec{x}) \in [\psi,\psi+
    \Delta \phi)$ follows from $\vec{x}$ being a uniformly distributed random
    variable. If $P(\phi)$ is the p.d.f.\ of $\phi$ then, from the definition
    of a p.d.f.,
\begin{equation}
\label{eq:prop_volume_relation}
\frac{V(\psi,\Delta \phi)}{V} = p(\phi;\psi \leq \phi + \Delta \phi) =
\int_{\psi}^{\psi + \Delta \phi} P(\phi) d\phi \ .
\end{equation}
Using the rectangle rule, the left hand side in
\eqref{eq:federer_coarea_integral} can be written
\begin{equation}
\label{eq:error_area_integral}
\int_{\psi}^{\psi + \Delta \phi} A(\phi) d\phi = {A(\psi) \Delta \phi} +
k_A(\Delta \phi) {\Delta \phi}^2 \ ,
\end{equation}
where $k_{A}(\Delta \phi)$ denotes the maximum value of ${d A(\phi)}{d\phi}$
in the domain $[\psi,\psi + \Delta \phi)$.  Using
\eqref{eq:federer_coarea_integral}, \eqref{eq:prop_volume_relation} and
\eqref{eq:error_area_integral}, we arrive at
\begin{equation}
\label{eq:coarea_arearatio_error1}
\frac{{A(\psi) \Delta \phi} + k_A(\Delta \phi) {\Delta \phi}^2}{V} =
     {\int_{\psi}^{\psi + \Delta \phi} P(\phi)
       d\phi}\ {\langle|\nabla\phi|\rangle_{V(\psi,\Delta \phi)}} \ .
\end{equation}
Dividing through by $\Delta \phi$ yields 
\begin{equation}
\label{eq:coarea_arearatio_error}
\frac{A(\psi)}{V}  = \frac{\int_{\psi}^{\psi + \Delta \phi} P(\phi)
  d\phi}{\Delta \phi} \ {\langle|\nabla\phi|\rangle_{V(\psi,\Delta \phi)}} -
\frac{k_A(\Delta \phi)}{V} {\Delta \phi} \ ,
\end{equation}
which is a first order approximation for the area to volume ratio.  
Finally, approximating the integral of the
p.d.f.\ with the rectangle rule results in
\begin{equation}
\label{eq:coarea_arearatio_withpdf}
\frac{A(\psi)}{V}  = \frac{P(\phi;\phi=\psi)
  \Delta \phi + k_p(\Delta \phi){\Delta \phi}^2}{\Delta \phi}
\ {\langle|\nabla\phi|\rangle_{V(\psi,\Delta \phi)}} - \frac{k_A(\Delta
  \phi)}{V} {\Delta \phi} \ ,
\end{equation}
where $k_{P}(\Delta \phi)$ is the maximum value of ${d P(\phi)}{d\phi}$ in the
domain $[\psi,\psi + \Delta \phi)$.  By comparing
\eqref{eq:coarea_arearatio_withpdf} and \eqref{eq:coarea_arearatio} we see
that, if the Monte-Carlo integration of the latter is converged, then it
provides a first-order approximation of $P(\phi;\phi=\psi)$. The integral in
\ref{eq:error_area_integral} can also be evaluated with a higher order
approximation. \citet[][Chapter 5, Figure 5.10]{shete19} evaluates the
integral using the 
midpoint rule, which is a second order method, and obtain second order
convergence with layer thickness. The definition of $P(\phi)$ in the turbulence
literature, e.g., \citep{pope00}, is as the limit of a histogram with bins
extending from $\phi$ to $\phi+\Delta\phi$.  To strictly preserve the
relationships discussed in this appendix we have used first-order
integration to evaluate isosurface areas in this paper. 

\section{Effect of Large Scales}  
Equation \eqref{eq:federer_coarea} shows that an isosurface area is a function
of $\nabla \phi$ and \citet{catrakis02} conclude that while large and small
length scales in a jet contribute to the area-volume ratio, the dominant
effect is at the small scales.  Therefore, we do not expect the area
calculations to be strongly affected by domain size.  Also, in conjunction
with figure \ref{fig:peclet} we assert that, while the equations and boundary
conditions describe a homogeneous flow, the simulations are not perfectly
homogeneous because the domain size is finite so that different isosurfaces
have slightly different areas; the error bars in figure \ref{fig:peclet}
indicate the extent to which the statistics of the large scales are not
resolved in each simulation.  In principle the large-scale statistics can be
improved by aggregating them over time, but it is
impractical to advance large simulations for sufficiently long times and doing
so would not improve the statistics for a single snapshot of the flow.
Therefore, in this appendix, we examine the effects on our conclusions of the large
scales and their statistical convergence in single snapshots in time. 

We compare here the isosurface areas for pairs of equivalent simulations.  The
first simulation in each pair is that used for the main body this paper and
denoted either R24 or R42.  The second simulation in each pair is
statistically the same but resolved in a domain eight times larger in each
dimension.  The small-scale resolution for both variants of R24 is the same
while for R42 the grid spacing in the large domain is twice that in the small
domain.  The ratio of the domain size to the integral length scale ${\cal
  L}/L$ is approximately 45 for the large domain size, which is significantly
larger than the value of 20 reported by \citet{debk98b} as being necessary for
decaying HIT to have sufficiently large-scale resolution to decay at the same
rate as grid turbulence in a wind tunnel.

The results from this test are tabulated in table \ref{tbl:tblA} for the
scalar fields with the highest and lowest Schmidt numbers.
In each case, the mean surface area for the large domain is within the range of
areas for the small domain, and the difference between the minimum and maximum
areas is smaller for the larger domain.  We conclude that the surface areas
are insensitive to the size of the domain and that the areas of all
isosurfaces approach the same value as the domain size increases.
\begin{table}
	\begin{center}
		\begin{tabular}{l l r r r r r r r r r}
&&  && \multicolumn{3}{c}{R24} &&   \multicolumn{3}{c}{R42} \\ 
Domain && Sc &&  mean & min & max && mean & min & max \\
\cline{1-1} 
\cline{3-3} 
\cline{5-7} 
\cline{9-11} 
Small && 0.1 && 3.82 & 3.32 & 4.32 &\hspace*{2em}& 4.16 & 3.84 & 5.00 \\Large && 0.1 && 3.91 & 3.89 & 3.93 && 4.25 & 4.20 & 4.40 \\\\
Small && 7.0 && 23.58 & 20.68 & 25.58 &\hspace*{2em}& 30.84 & 28.68 & 33.14 \\Large && 7.0 && 24.17 & 23.88 & 24.42 && 30.91 & 30.54 & 31.15 \\\end{tabular}
	\end{center}
	\caption{The mean, minimum, and maximum isosurface areas in equivalent
          simulations run in differing domain sizes.  All areas are per unit
          volume times the volume of the smaller domain.
          \label{tbl:tblA}}
\end{table}

\bibliographystyle{jfm}
\bibliography{bib}

\end{document}